%% file: moriond.tex
\documentclass[11pt]{article}
\usepackage{moriond,epsfig}

\usepackage[T1]{fontenc}        
\usepackage[french]{babel}      
\include{babarsym}
\include{def}

\def\Journal#1#2#3#4{{#1} {\bf #2}, #3 (#4)}

\def\PLB{{\em Phys. Lett.}  B}
\def\PRL{\em Phys. Rev. Lett.}
\def\PRD{{\em Phys. Rev.} D}

\def\EPJ{{\em Eur. Phys. J.} C}
\def\RMP{{\em  Rev. Mod. Phys.}}

\def\be{\begin{equation}}
\def\ee{\end{equation}}
\def\bea{\begin{eqnarray}}
\def\eea{\end{eqnarray}}

\begin{document}
\vspace*{4cm}
\title{DEPUZZLING $B\to K\pi$ : CONSTRAINTS ON THE UNITARITY TRIANGLE FROM $B,B_s\to \pi\pi,K\pi,KK$ DECAYS IN THE $SU(3)$ LIMIT}

\author{J. MALCL\`ES~\footnote{
    Talk presented by J. MALCL\`ES based on work in collaboration with J. CHARLES and J. OCARIZ~\cite{TheSU3Pap}. 
    We thank A. H\"OCKER for important collaboration at the early stage of this work. }}

\address{Laboratoire de Physique Nucl\'eaire et des Hautes Energies,  IN2P3 - CNRS Universit\'es Paris VI et VII,
 4 place Jussieu, Tour 33, Rez-de-chauss\'ee ,  75252 Paris Cedex 05, France}

\maketitle\abstracts{
 Constraining CKM parameters from charmless hadronic $B$ decays 
 requires methods for addressing the hadronic uncertainties.
A complete technique is presented, using  relations between amplitudes in the
$B,B_s\to \pi\pi, K\pi, KK$ system obtained in the exact $SU(3)$ symmetry limit,
without having to neglect annihilation/exchange 
topologies.
Naive $SU(3)$-breaking effects in the decay amplitudes are taken into account, through the
inclusion of 
$\pi$ and $K$ decay constants in the normalisations and conservative theoretical errors.
Already with the limited set of observables currently available, significant constraints on the CKM 
parameters are obtained. 
Also, subsets of observables are shown to bring
 non trivial constraints on the CKM angles $\alpha$ and $\beta$, in agreement with analytical solutions that we derive.
Finally, the future  physics potential of this method is estimated, in light of the increased
precision of measurements from the current B-factories, and the inclusion of several  
new observables from $B_s$ decays expected to be provided by the LHC experiments.
}

\section{Introduction}
\input{intro}

\section{Formalism}
\label{formalism}
	\subsection{Model-independent parameterization in the SU(3) limit}
	\input{param}

	\subsection{Electroweak penguins from $Q_{9,10}$ dominance}
	\input{pew}

	\subsection{SU(3) breaking}
	\input{su3break}

\section{Some observables subsets}
\label{sec:subsets}

 Within this framework, one can first reduce the number of unknowns by considering subsystems of observables 
 constraining the angles $\alpha$ and $\beta$ separately. 
 These subsystems are of great interest because they dominate the constraints in the \rhoeta\ plane 
 and they can be solved analytically. For the sake of simplicity, we will give the analytical solutions 
 in the case of vanishing annihilation and exchange topologies. Note that analytical solutions do still exist
 without this hypothesis, and that the numerical results do not neglect these contributions.

	\subsection{The ``$\alpha$'' subsystem}
	\input{alpha}

	\subsection{The ``$\beta$'' subsystem}
	\input{beta}

\section{Input data, parameter counting and statistical approach}
\label{sec:inputs}
\input{input}

\section{Fit results}
\label{sec:results}

	\subsection{The ``$\alpha$'' and ``$\beta$'' subsystems}

\input{fitalphabeta}

	\subsection{Joint and full systems}
	\input{rhoeta}
		
	\subsection{Future physics potential}
	\label{2008}
	\input{su308}

\section{Conclusion and outlook}

Already with the limited set of observables currently available, significant constraints on the CKM 
parameters are obtained. 
Also, observables from the  $B^0\to\pi^+\pi^-$, $B^0\to K^+\pi^-$, and $B^0\to K^+K^-$ subsystem 
alone are shown to bring strong constraints on the CKM angle $\alpha$. A similar constraint on $\beta$ is obtained from
 the subsystem  $B^0\to\pi^0\pi^0$, $B^0\to K^0\pi^0$, and $B^0\to K^+K^-$.  
The full constraint on the apex on the Unitarity Triangle can already
be compared with the standard CKM global fit.
 In the future, this framework alone will be able to determine the 
   Unitarity Triangle with an accuracy comparable to the current CKM fit, and could be used to constraint SU(3) 
 breaking or New Physics parameters.

\end{document}

%% file: def.tex
\newcommand{\bei}{\begin{itemize}}
\newcommand{\eei}{\end{itemize}}
\newcommand{\beq}{\begin{equation}}
\newcommand{\eeq}{\end{equation}}
\newcommand{\beqn}{\begin{eqnarray}}
\newcommand{\eeqn}{\end{eqnarray}}
\newcommand{\beqns}{\begin{eqnarray*}}
\newcommand{\eeqns}{\end{eqnarray*}}

\newcommand\Amptpbar{\kern 0.18em\overline{\kern -0.18em {\cal A}}_{3\pi}}

\newcommand\Amptpbarkappa{\kern 0.18em\overline{\kern -0.18em A}^{\kappa}{}}
\newcommand\Amptpbarsigma{\kern 0.18em\overline{\kern -0.18em A}^{\sigma}{}}

\newcommand\rhoeta{(\rhobar,\etabar)}

\newcommand\Nbpm{{\kern 0.18em\overline{\kern -0.18em N}}^{+-}}
\newcommand\Nbmp{{\kern 0.18em\overline{\kern -0.18em N}}^{-+}}

\newcommand\BRpmb{{\cal \kern 0.18em\overline{\kern -0.18em  B}}{}_{\rho\pi}^{+-}}
\newcommand\BRmpb{{\cal \kern 0.18em\overline{\kern -0.18em  B}}{}_{\rho\pi}^{-+}}

\newcommand\BRipmb{{\cal \kern 0.18em\overline{\kern -0.18em  B}}{}_{\rho^+\pi^-}}
\newcommand\BRimpb{{\cal \kern 0.18em\overline{\kern -0.18em  B}}{}_{\rho^-\pi^+}}

\newcommand\Abar{\kern 0.18em\overline{\kern -0.18em A}{}}

%% file: intro.tex
 Constraining CKM parameters from charmless hadronic $B$ decays requires methods for addressing the hadronic uncertainties.
 A common method consists in considering symmetries to relate different decay amplitudes 
 and eliminate hadronic unknowns. SU(2) symmetry is well understood and largely used
  to get constraints on the Unitarity Triangle from charmless two-body hadronic $B$ decays.
 In the $\pi \pi$ system, it allows to determine the angle $\alpha$ up to an eightfold ambiguity, 
 whereas in the $K \pi$ system, it requires additional hypotheses to be predictive.
 In both cases, the derived constraints remain weak with the current errors and one can wonder how to use
 all the available inputs in a more efficient manner. 
 Although errors remain large, theoretical calculations as SCET
 or QCD factorization can be considered, as it has been discussed previously~\cite{myref,Bauer}.
  Here, we propose a new data-driven technique, using relations between amplitudes in the
 $B,B_s\to \pi\pi, K\pi, KK$ system obtained in the $SU(3)$ symmetry limit. 
 Use of approximate SU(3) symmetry~\cite{Silva} for those modes has received considerable attention in the 
recent literature~\cite{BFRS1,BFRS2,Wu,ThePap}. In this paper, the exact $SU(3)$ limit is considered,
  without additional hypotheses such as the neglect of annihilation/exchange 
topologies. Naive $SU(3)$-breaking effects in the decay amplitudes are taken into account, through the
inclusion of $\pi$ and $K$ decay constants in the normalisations and conservative theoretical errors.

 The first section is devoted to the formalism: the decay amplitudes under $SU(3)$ are expressed, 
 electroweak penguin amplitudes are related to other amplitudes in a model-independent way,
  and the SU(3) breaking parameterization is described. 
 In section~\ref{sec:subsets}, two analytically solvable subsystems of observables are introduced mainly 
 constraining the CKM angles $\alpha$
 and $\beta$.
 The inputs, the parameter counting and the statistical approch are briefly discussed in section~\ref{sec:inputs}.
  Finally, numerical results are given in section~\ref{sec:results}, where the future 
 physics potential of this method is also estimated, in light of the increased
precision of measurements from the current B-factories, and the inclusion of several  
new observables from $B_s$ decays expected to be provided by the LHC experiments.
This work will be described in further details in an upcoming publication~\cite{TheSU3Pap}.

%% file: param.tex
Benefiting from the unitarity of the CKM matrix, one can provide a phenomenological description of any decay amplitude 
in terms of CKM matrix elements and two complex hadronic amplitudes; namely for the $B^{0} \rightarrow \Kp \pim$ decay
 amplitude: $ A(\Kp \pim) = V_{us}V_{ub}^* \,  T^{+-} + V_{ts}V_{tb}^* \,  P$.
Owing to SU(3) invariance of strong interaction, the amplitudes of various $B,B_s\to \pi\pi,K\pi,KK$ decays are related to each other
  through 16 complex independent equations, and thus the system can be described \textit{via} only 10 hadronic amplitudes,
  i.e. 19 real physical parameters, as follows:
\beqn
A(B_d \to \Kp \pim)      &=&   V_{us}V_{ub}^* \,  T^{+-} + V_{ts}V_{tb}^* \,  P~,\nonumber \\
A(B_u \to \Kz \pip)      &=&   V_{us}V_{ub}^* \,  N^{0+} + V_{ts}V_{tb}^* \,  (-P+P^{EW}_C),\nonumber \\
A(B_u \to \Kp \piz)      &=&   V_{us}V_{ub}^* \,  (T^{+-}+T^{00}-N^{0+}) + V_{ts}V_{tb}^* \,  (P+P^{EW}-P^{EW}_C)~,\nonumber \\
A(B_d \to \Kz \piz)      &=&   V_{us}V_{ub}^* \,  T^{00} + V_{ts}V_{tb}^* \,  (-P+P^{EW})~,\nonumber \\
A(B_d \to \pip \pim)     &=&   V_{ud}V_{ub}^* \,  (T^{+-}+\Delta T) +  V_{td}V_{tb}^* \,  (P+PA)~,\nonumber \\
A(B_d \to \piz \piz)     &=&   V_{ud}V_{ub}^* \,  (T^{00}-\Delta T) + V_{td}V_{tb}^* \,  (-P-PA+P^{EW})~,\nonumber \\
A(B_u \to \pip \piz)     &=&   V_{ud}V_{ub}^* \,  (T^{+-}+ T^{00}) + V_{td}V_{tb}^* \,  P^{EW}~,\nonumber \\
A(B_d \to \Kp \Km)       &=&   V_{ud}V_{ub}^* \,  \Delta T + V_{td}V_{tb}^* \,  PA~,\nonumber \\
A(B_d \to \Kz \bar{\Kz}) &=&   V_{ud}V_{ub}^* \,  \Delta P + V_{td}V_{tb}^* \,  (-P-PA+ P^{EW}_C-\frac{1}{2} P^{EW}_{K\bar{K}})~,\nonumber \\
A(B_u \to \Kp \bar{\Kz}) &=&   V_{ud}V_{ub}^* \,  N^{0+} + V_{td}V_{tb}^* \,  (-P+P^{EW}_C)~,\nonumber \\
A(B_s \to\Kp \Km)     &=&   V_{us}V_{ub}^* \,  (T^{+-}+\Delta T) + V_{ts}V_{tb}^* \,  (P+PA)~,\nonumber \\
A(B_s \to\Kp \pim)    &=&   V_{ud}V_{ub}^* \,  T^{+-} +  V_{td}V_{tb}^* \,  P~,\nonumber \\
A(B_s \to\pip \pim)   &=&   V_{us}V_{ub}^* \,  \Delta T + V_{ts}V_{tb}^* \,  PA~.
\label{eq:SU3}
\eeqn         
These equations are perfectly exact in the SU(3) limit.

%% file: pew.tex
One can relate the electroweak penguins amplitudes $P^\mathrm{EW}$, 
 $P^\mathrm{EW}_\mathrm{C}$ and $P^{EW}_{K\bar{K}}$ to the other amplitudes in a model-independent way~\cite{BFPEW,NRPew} in the SU(3) limit
 making use of Fierz transforms and benefiting from the dominance of the operator $c_9 O_9 + c_{10} O_{10}$ with respect to $c_7 O_7 + c_8 O_8$:
\beqn
\label{PEWSU3}
P^\mathrm{EW} &=&
R^+ \left(T^{+-} + T^{00}\right)~,\nonumber\\
P^\mathrm{EW}_\mathrm{C} &=&
\frac{R^+}{2} \left(T^{+-} + T^{00} + N^{0+} - \Delta T - \Delta P\right) -
\frac{R^-}{2} \left(T^{+-} - T^{00} + N^{0+} + \Delta T + \Delta P\right)~,\nonumber\\
P^\mathrm{EW}_\mathrm{K\bar{K}} &=&
\frac{R^+}{2} \left(N^{0+} - \Delta T - \Delta P\right)~.
\eeqn
In the above equations $R^+$ and $R^-$ are constants given by
\beq
\label{R+R-}
        R^+     =     -\frac{3}{2}\frac{c_9+c_{10}}{c_1+c_2}
                 \;=\;  +(1.35 \pm 0.12)~10^{-2}~,~~
        R^-     =     -\frac{3}{2}\frac{c_9-c_{10}}{c_1-c_2}
                 \;=\;  +(1.35 \pm 0.13)~10^{-2}~.
\eeq
The theoretical error on the numerical evaluation of this ratio has
been estimated from the residual scale and scheme dependence of the
Wilson coefficients~\cite{bbl}. 

%% file: su3break.tex
SU(3) flavor symmetry is only approximately realized in nature and one
may expect violations up to $30\%$ at the amplitude level.
For example, within factorization the relative size of SU(3) symmetry
breaking is expected to be $(f_{K}-f_{\pi})/f_{K}$, where $f_K$ and $f_\pi$
are the pion and kaon decay constants, respectively. Dominant factorizable SU(3) breaking effects~\cite{kho}
 are taken into account \textit{via} the normalization of  $B \to K \pi$, $B \to K \bar{K}$ and $B_s \to K \pi$ amplitudes, 
 with regard to the $B \to \pi \pi$ one. The normalization factors are respectively:
\beqn
N_{K \pi} &=&  \frac{f_K}{f_\pi}= 1.22 \pm 0.22~,\nonumber\\
N_{K \bar{K}}&=&  \left(\frac{f_K}{f_\pi}\right)^2  \frac{f_{B_s}}{f_{B_d}} = 1.81 \pm 0.34 ~,\nonumber\\
N^s_{K \pi} &=&  \frac{f_K}{f_\pi} \frac{f_{B_s}}{f_{B_d}} = 1.48 \pm 0.28 ~,
\label{norm}
\eeqn 
where the theoretical uncertainty of $22\%$ is calculated taking the error on $\frac{f_K}{f_\pi}$ to be its deviation from one. 
 Remaining SU(3) breaking effects are neglected:  residual factorizable SU(3)
breaking does not exceed a few percents, while non factorizable SU(3) breaking sources, being unconstrained by both theoretical
 and experimental arguments for the moment, are assumed to play no important role.


%% file: alpha.tex
	Let us first consider all the observables related to  $B^0\to\pi^+\pi^-$, $B^0\to K^+\pi^-$, and $B^0\to K^+K^-$ decays 
	and call this subsystem ``$\alpha$''.
	In this case, the system reduces to 4 complex hadronic unknowns: $T^{+-}$, $P$, $\Delta T$ and $PA$.
	Neglecting annihilation and exchange topologies ($\Delta T=PA=0$), the system can be described by two equations,
	 in the $V_{us}V_{ub}^* ,  V_{cs}V_{cb}^*$ basis:
	\beq
	A(B \to \Kp \pim) = t e^{i \gamma} - p ~, \nonumber 
	A(B \to \pip \pim) =  \lambda t e^{i \gamma} - \frac{1}{\lambda} p  ~, \nonumber \\
	\eeq
	solving to the analytical solution:
	\beq
	\sqrt{1-C_{\pi \pi}^{+- \,2}} |\mathcal{D}| \cos(2\alpha-2\alpha_{\mathrm{eff}} - \epsilon) = (1+\lambda^2)^2 -2 \lambda^2 sin^2(\gamma) 
	 \left(1 + \frac{{\cal B}_{K \pi}^{+-}}{{\cal B}_{\pi \pi}^{+-}}\right)~,
         \label{eq:su3alpha}
	\eeq
	with ${\cal B}_{K \pi}^{+-} C_{K \pi}^{+-} + {\cal B}_{\pi \pi}^{+-} C_{\pi \pi}^{+-} =0$ and $\mathcal{D} \equiv |\mathcal{D}| e^{i \epsilon} = (1+\lambda^2)\, (1+\lambda^2 e^{i \gamma})$. Thus, this subsystem mainly measures the angle $\alpha$, with a $\lambda^2$ suppressed dependence on the angle 
	$\gamma$ ($\lambda=\sin(\theta_{\rm Cabibbo}) \sim 0.23$).

%% file: beta.tex
Let us now consider the subsystem of observables related to $B^0\to\pi^0\pi^0$, $B^0\to K^0\pi^0$, and $B^0\to K^+K^-$ decays.
In the same manner than for the ``$\alpha$'' subsystem, the hadronic unknowns are reduced to four complex quantities $T^{00}$, $-P+P^{EW}$, $\Delta T$ et $PA$. 
 Neglecting annihilation and exchange topologies, it solves to:
	\beq
	\sqrt{1-C_{K \pi}^{00 \, 2}} |\mathcal{D}| \cos(2\beta-2\beta_{\mathrm{eff}} + \epsilon) = (1+\lambda^2)^2 -2 \lambda^2 sin^2(\gamma) 
	\left(1 + \frac{{\cal B}_{\pi \pi}^{00}}{{\cal B}_{K \pi}^{00}}\right)~,
        \label{eq:su3beta}
	\eeq
	with ${\cal B}_{K \pi}^{00} C_{K \pi}^{00} + {\cal B}_{\pi \pi}^{00} C_{\pi \pi}^{00} =0$.
	This subsystem, called ``$\beta$'' in the following, mainly measures the angle $\beta$, with a $\lambda^2$ suppressed dependence on the angle 
	$\gamma$.

%% file: input.tex
 The inputs used are world averages from HFAG~\cite{HFAG} at EPS 2005,
 using available results from \babar, Belle, CLEO and CDF.
 We have used ten branching ratios, eight CP asymmetries, and the two CP parameters 
 $S^{+-}_{\pi \pi}$ and $S^{00}_{K_S \pi}$; i.e. 20 independent observables 
 in total for the $B_d$ and $B_u$ amplitudes. In addition, we have taken into account six ratios of
  branching ratios measured by CDF, four of which are the only observables available for $B_s$ decays.

 For the ``$\alpha$'' and ``$\beta$'' subsystems, we have respectively 8 and 6 measurements available, 
 both corresponding to 6 independent observables,
 for a total of 7 real hadronic unknowns for each system.
 For the full system, we already have in total 21 independent observables for 13 real hadronic unknowns. 
 There are up to 38 independent observables related to this system
  making it a promising tool to constraint the
Unitarity Triangle in the future.

We use classical frequentist statistics (minimum $\chi^2$) to obtain
the constraints on the parameters. All unknown parameters are left free
to vary in the fit without contributing to the $\chi^2$. As for the parameters
that come with a theoretical uncertainty (namely $R^{\pm}$, and the
normalization factors in~(\ref{norm})), we use the Rfit~\cite{TheBible} approach.


%% file: fitalphabeta.tex
\begin{figure}\begin{minipage}{.49\linewidth}
\psfig{figure=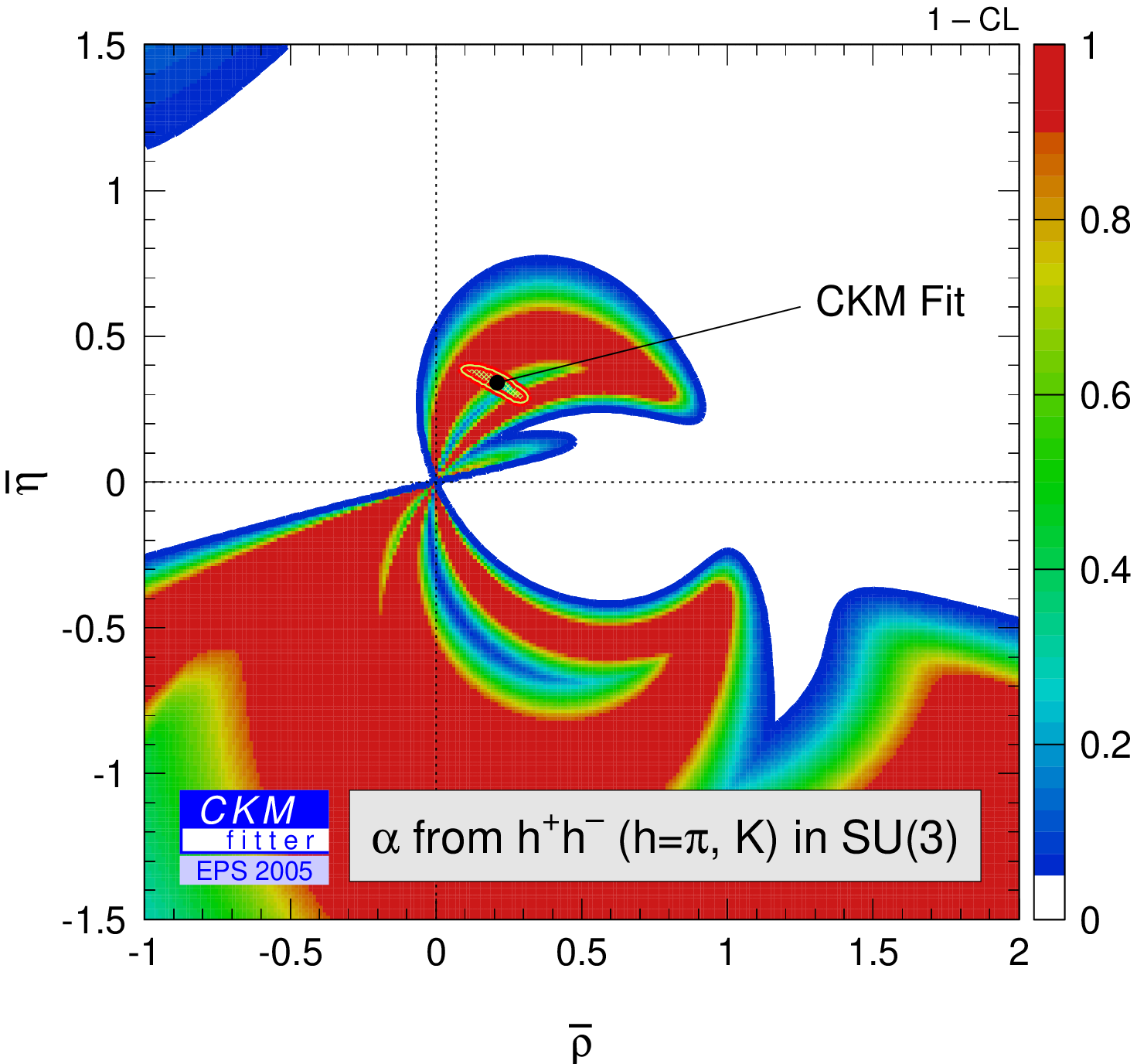,height=2.6in}\end{minipage}\begin{minipage}{.49\linewidth}
\psfig{figure=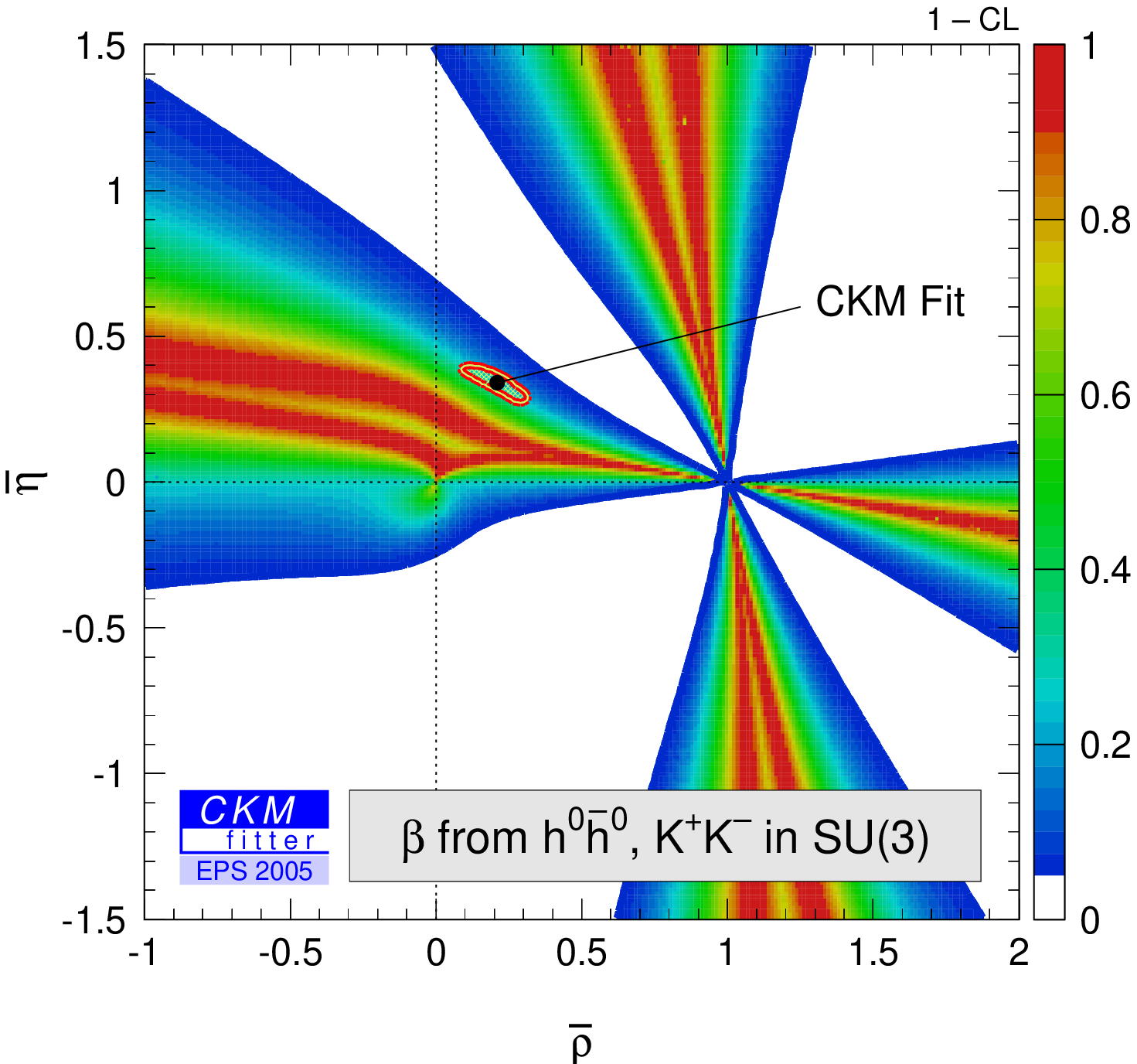,height=2.6in}\end{minipage}
\caption{Constraints in the \rhoeta\ plane from $B\to \pip\pim, \Kp \pim, \Kp \Km$ modes,
 i.e. from the ``$\alpha$'' subsystem of inputs (left) and from  $B\to \piz\piz, \Kz \piz, \Kp \Km$ modes,
 i.e. from the ``$\beta$'' subsystem of inputs (right).
\label{fig:alpha}}
\end{figure}

Constraints in the \rhoeta\ plane for the ``$\alpha$'' and ``$\beta$'' subsystems are shown on figure~\ref{fig:alpha}.
The respective dependence on the angles $\alpha$ and $\beta$ is clearly visible, whereas the dependences on the angle $\gamma$
 create small structures breaking the symmetry of the constraints. Both results are in good agreement with the superimposed standard CKM fit.

%% file: rhoeta.tex
\begin{figure}\begin{minipage}{.49\linewidth}
\psfig{figure=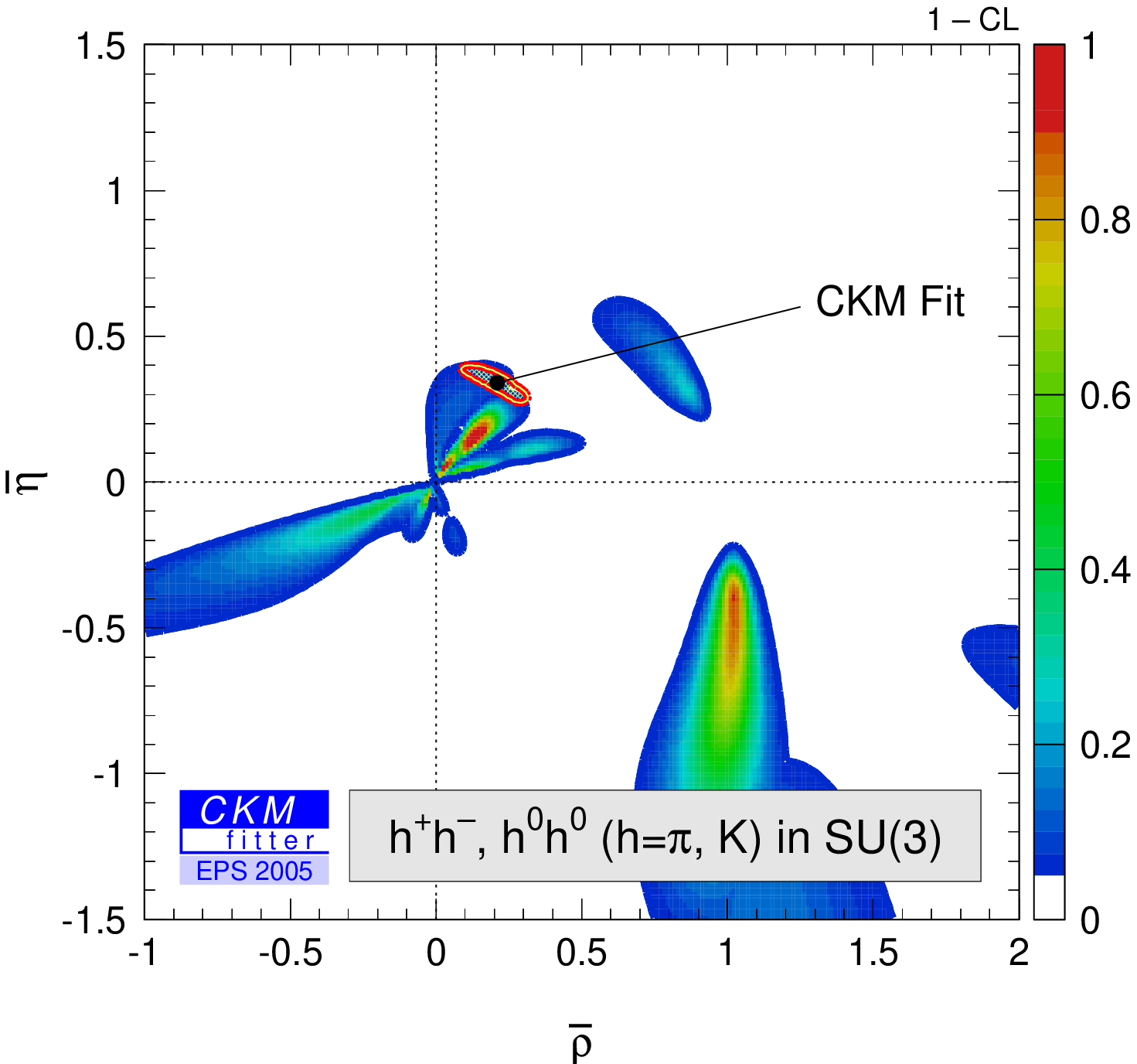,height=2.6in}\end{minipage}
\begin{minipage}{.49\linewidth}
\psfig{figure=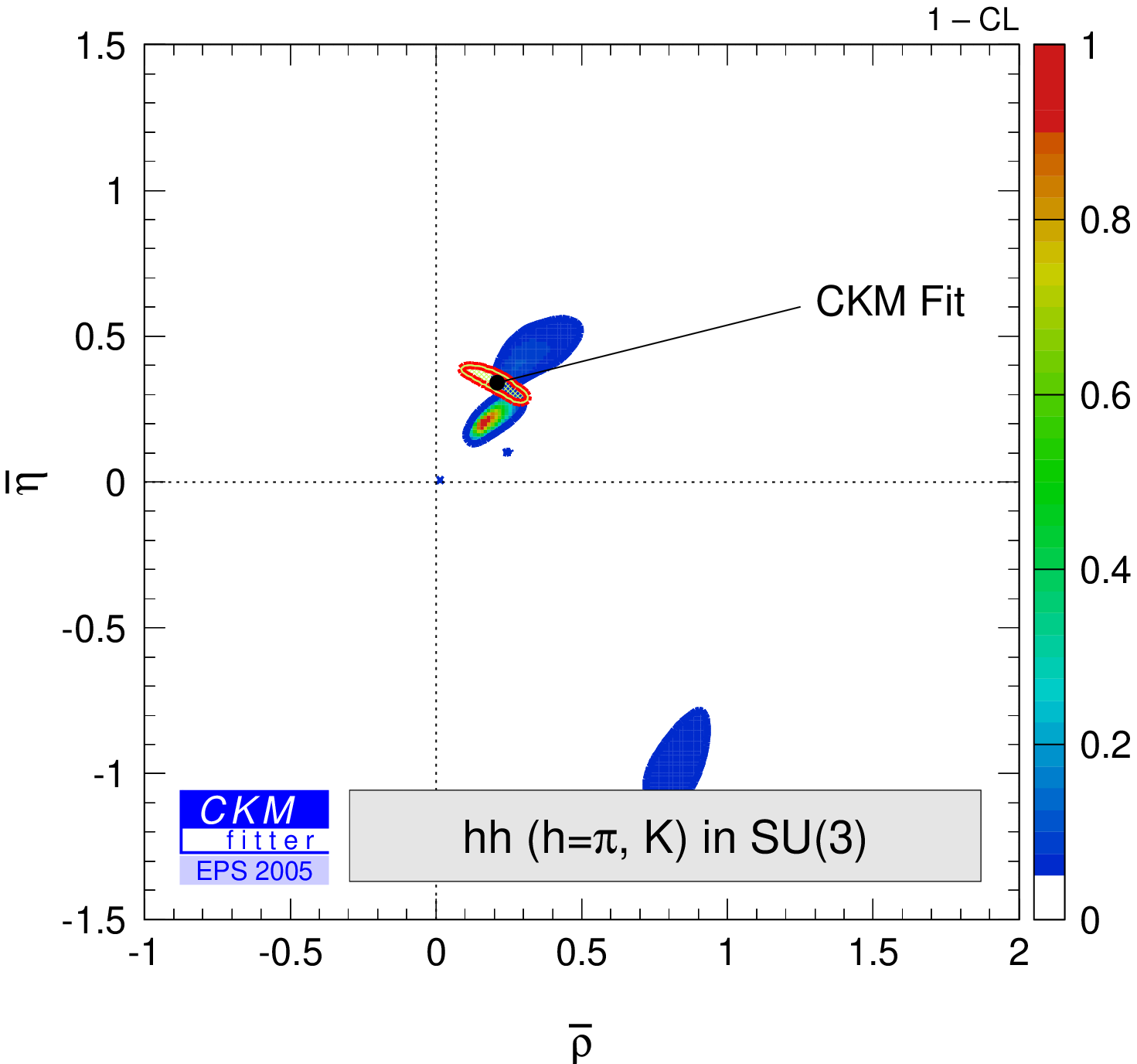,height=2.6in}\end{minipage}
\begin{minipage}{.49\linewidth}
\psfig{figure=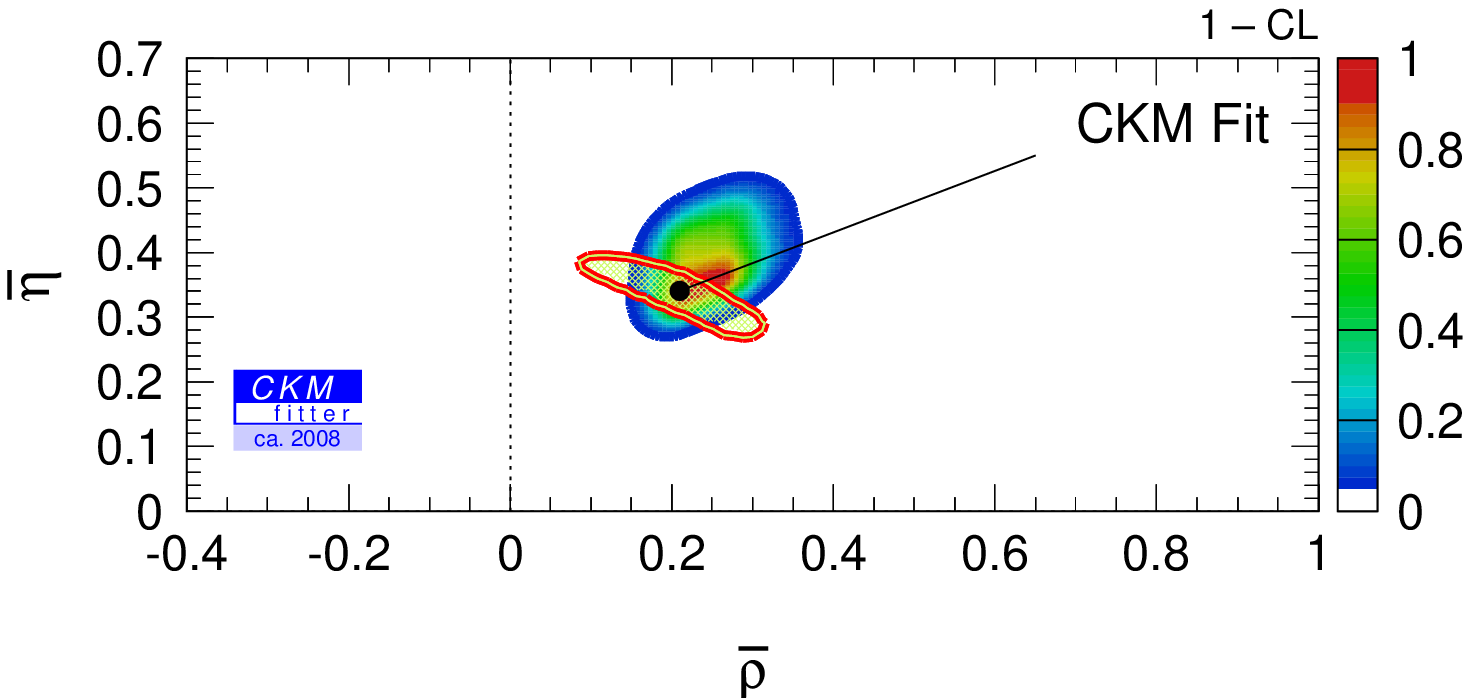,height=1.3in}\end{minipage}
\begin{minipage}{.49\linewidth}
\psfig{figure=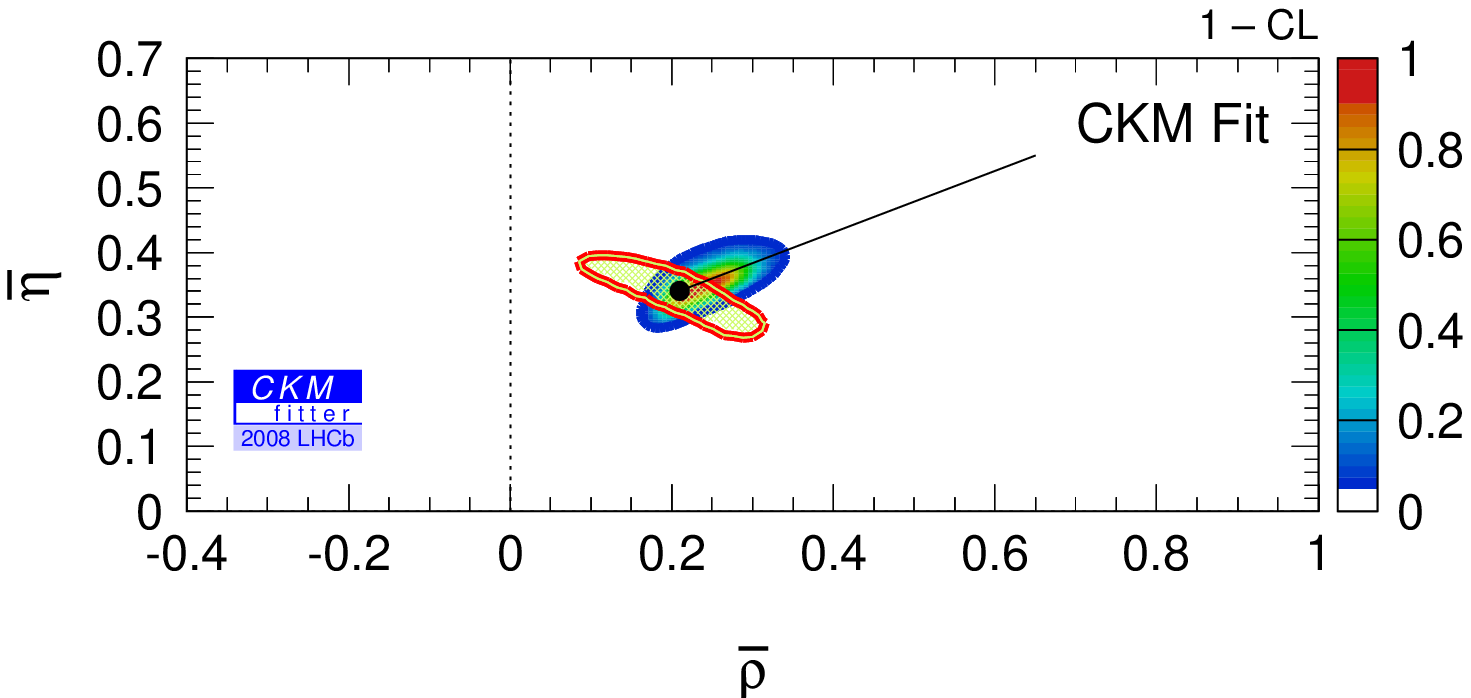,height=1.3in}\end{minipage}
\caption{Top: constraints in the \rhoeta\ plane from the joint ``$\alpha$'' and ``$\beta$'' subsystems (left) and for the full system (right).
Bottom: constraints in the \rhoeta\ plane for the full system with errors expected in 2008 including two B factories and CDF (left) and adding three inputs from LHCb (right).
\label{fig:full}}
\end{figure}

 Correlations between the two subsystems arise from two sources: the common mode $B_d \to \Kp \Km$
 and the tree-dependent parameterization of the electroweak penguins amplitude  $P^{EW}$. If $P^{EW}$ was totally free, 
 $-P + P^{EW}$ could be identified with an independent free amplitude in the ``$\beta$'' subsystem, and in case of vanishing 
 annihilation and exchange topologies ($\Delta T=PA=0$), the two subsystems would be completely uncorrelated.
The constraints obtained joining the systems ``$\alpha$'' and ``$\beta$'' are shown on figure~\ref{fig:full} (top left).
 They are stronger than the naive product of separated contraints assuming the absence of correlations.
 This effect comes mainly from the expression of the electroweak penguins amplitude $P^{EW}$. We also find that the fit marginally
prefers non standard valeurs for $P^{EW}$, in agreement with what was argued by the authors of~\cite{Gronau,BFRS1,BFRS2}.

The constraints obtained with the full system and all available inputs are also shown on figure~\ref{fig:full} (top right).	
The additional inputs allow to eliminate mirror solutions and get strong constraints, in reasonable agreement with the standard CKM fit. 

%% file: su308.tex
To estimate the future physics potential of this method, we have performed 
a tentative analysis using the  errors expected in 2008. Central values have been chosen to be 
 the best fit values for the current set of inputs in the full system framework.
The two bottom plots of figure~\ref{fig:full} show the induced constraints from a
 closest view in the \rhoeta\ plane. On the left, two B factories with $1000fb^{-1}$
  each have been considered~\cite{PhysReach}; and on the right,
 three inputs from LHCb have been added~\cite{TDRLHCB} ($C^{+-}_{B_s \to KK}$, $S^{+-}_{B_s \to KK}$ and ${\cal A}^{+-}_{B_s \to K \pi}$). 
 Including LHCb results, the contraints are found to be competitive with 
 the current CKM fit demonstrating the predictive power of this framework.